# Elucidating Pathfinding Elements from the Kubi Gold Mine in Ghana

**Gabriel K. Nzulu\*, Babak Bakhit, Hans Högberg, Lars Hultman and Martin Magnuson**
Department of Physics, Chemistry and Biology (IFM) Linköping University, SE-58183, Linköping Sweden
\*Correspondence: Gold Corporation, No. 1, Yapei Link, Airport Residential Area, P.O. Box 9311, Airport, Accra, Ghana

**Abstract:** X-ray photoelectron spectroscopy (XPS) and energy dispersive X-ray spectroscopy (EDX) are applied to investigate the properties of fine-grained concentrate on artisanal small-scale gold mining samples from the Kubi Gold Project of the Asante Gold Corporation near Dunwka-on-Offin in the Central Region of Ghana. Both techniques show that the Au-containing residual sediments are dominated by the host elements Fe, Ag, Al, N, O, Si, Hg, and Ti that either form alloys with gold or inherent elements in the sediments. For comparison, a bulk nugget sample mainly consisting of Au forms an *electrum i.e.*, a solid solution with Ag. Untreated (impure) sediments, fine-grained Au concentrate, coarse-grained Au concentrate, and processed ore (Au bulk/nugget) samples were found to contain cluster of O, C, N, and Ag with Au concentrations significantly lower than that of the related elements. This finding can be attributed to primary geochemical dispersion, which evolved from crystallization of magma and hydrothermal liquids as well as migration of metasomatic elements and rapid rate of chemical weathering of lateralization in secondary processes. The results indicate that Si and Ag are strongly concomitants with Au because of their eutectic characteristics N, C, and O follow alongside because of their affinity to Si. These non-noble elements thus act as pathfinders of Au ores in the exploration area. The paper further discusses relationships between gold and sediments of auriferous lodes as key to determine indicator minerals of gold in mining sites.

**Introduction**
Gold mineralization in the Kubi Gold Project of the Asante Gold Corporation near Dunwka-on-Offin in the Central Region of Ghana is mainly controlled by the silicate type of garnet mineral ($A_3B_2(SiO_4)_3$), where $A$ site represents the divalent cations (Ca, Mg, Fe, Mn)$^{2+}$ and the $B$ site has trivalent cations in either an octahedral or tetrahedral milieu with $(SiO_4)^{4-}$ occupying the tetrahedral sites [1,2]. Sulfide group of minerals like pyrite/marcasite ($FeS_2$), arsenopyrite (FeAsS), pyrrhotite ($Fe_{(1-x)}S$) and chalcopyrite ($CufeS_2$) are other important pathfinder minerals in the location of the Au ore and help understand the geology of Kubi.
Traditionally, coarse-grained concentrates are used in gold production in most large-scale mining, whilst the fine-grained powder sediments are considered as part of mining waste (tailings). Artisanal and small-scale gold mining, without the use of mercury also depends on crushing or milling Au particles into similar grain sizes, which can be easy separated using gravity concentration methods *i.e.*, panning, sluicing, shaking tables, spiral concentration, vortex concentration, and centrifuges [3]. It will therefore be economically viable to extract gold and other minerals from fine-grained materials under realistic conditions.
Indicator and pathfinder theories are deployed by geologists and explorers to gain information on the location of ore deposits, through methods such as; gold grain morphology (changes in Au surface shape via weathering and erosion as sediments are transported from far distances), gold grain inclusion (presence of other minerals in gold grains are used to provide information about the deposit type and mineralization associated), composition studies (to gather and examine mineral samples for information on distances and direction of transport) and geochemistry (identifying and analyzing minerals like silver, platinum, palladium, copper, lead, iron, telluride), to test clues and patterns to locate ore-bodies for gold [4,5].
To date, researchers have tried to understand the nature and properties of the various elements associated with gold as well as their chemical bonds and reaction mechanisms in relation to the gold deposition and agglomeration [6-9]. Several studies indicate that during ore deposition in hydrothermal systems, there are surface reductions or adsorption or both in gold precipitation [10-13].
From a geological perspective, it is important to study the diffusion, flow mechanism, and chemical bonding of host (indicator) minerals to understand the dynamics and evolution of Earth and other terrestrial planets. These features can influence the mantle convection, Earth processes (erosion, weathering, landslide, volcanic eruption, earthquake, etc.), and also give information on the Earth's thermo-chemical structure. This will enable to understand the geodynamics of the Earth, rheological properties (material deformation), kinetics of phase transformation, and thermal conductivity of minerals in the earth.

Gold-associated minerals such as sulfide group $FeS_2$, FeAsS, $CuFeS_2$, $Fe_{(1-x)}S$, and galena (PbS) make the tiny particles of gold invisible or to be attached to the metallic sulfide lattices [14-18]. Pyrite is one of the most common minerals that conglomerate gold in various ore deposits and is also part of the gold-bearing minerals (as free milling or refractory) in the Au exploration areas alongside garnet and the sulfide group [6, 7, 14].

Previously, X-ray photoelectron spectroscopy (XPS), scanning electron microscopy (SEM), electrochemical, and other techniques have been deployed in the study of gold deposition from aqueous solutions on pyrite and other gold-bearing minerals [7,8,16]. Pathfinder minerals of Au such as chalcopyrite and pyrite minerals in their natural states were studied using cyclic voltammetry, electrochemical impedance spectroscopy, XPS, and SEM, which also included microanalysis (SEM/EDX). In both minerals, oxygen containing $Fe^{3+}$ and $Cu^{2+}$ were found and also the measurements showed alteration in both minerals at different potential values [19]. Harmer *et al.* [20-23] studied the rate of leaching of Cu and Fe in chalcopyrite when percolated in perchloric acid and found the leach rate to be the same after longer period of time with the final analysis based on surface speciation and oxidation steps in the release of Cu and Fe into solutions and the polymerization from mono-sulfide to polysulfide. Wang *et al.* [24], used high-resolution transmission electron microscopy and XPS to characterize the morphological change in Au nano-rods (NRs) to understand the concentration of mercury from artisanal small-scale mining. It was found that the low concentration of mercury that formed and got deposited on the gold surface could be reversibly removed by an electrochemical stripping process without causing any change in terms of size or shape to of Au NRs. However, at high concentration, Hg did not only deposit onto the gold surface, but also got into the interior of Au NRs and transformed them into an irreversible gold nanospheres (Au NSs) due to the amalgamation between Au with Hg. Since the assembling of Au with Hg by small-scale artisanal miners has significant negative impact on the health of individuals and the environment, there is a need for a more environmentally sustainable separation mechanism.

XPS has been a contributory technique [25] in the study of products of oxidation [26] formed on sulfide minerals and with regards to the use of synchrotron radiation [27]; it is an added advantage in explaining mineral surface chemistry (mineral geochemistry) [28, 30], predominantly the primary surface conditions of materials formed by deformations and cracks [30-32]. The main advantage of XPS and resonant photoemission using synchrotron radiation in mineralogy is the ability to define comprehensive surface sensitive information from high-resolution spectra for samples of macroscopic sizes of different mineral phases. The distribution of oxidation products provides heterogeneity of mineral surfaces and because most gold bearing sulfide minerals contain other minerals and impurities with varying texture, it will be of interest to reveal the individual elements by XPS, which can then be assigned to the original host minerals. It is therefore essential to identify the existence of the different types of elements on both the bulk and impure sample surfaces using high resolution XPS and the distribution of those elements across the sample surface [33-35].

Mikhlin *et al.* [36] used XPS, AFM, and scanning tunnelling microcopy/scanning tunnelling spectroscopy (STM/STs) to characterize arsenopyrite samples oxidized in Au(III) chloride solutions at ambient temperatures to explain the moderation of oxidation in chloride solutions. Murphy and Strongin [37] conducted surface reactivity studies on pyrite ($FeS_2$) and pyrrhotite ($Fe_xS_{1-x}$) using different experimental techniques such as vacuum type experiments where electron and photon spectroscopies were applied and further analyzed microscopically using infra-red spectroscopy. XPS and X-ray absorption spectroscopy techniques were also used to investigate the structure of the original pyrite and sulfide surfaces [37, 38]. Acres *et al.*, [39] applied synchrotron-based XPS, near-edge X-ray Absorption Florescence Spectroscopy, and time of flight secondary ion mass spectrometry studies on smooth and rough fracture surfaces of chalcopyrite and pyrite samples. The results showed an increase in the chalcopyrite surface roughness and an intensification in the formation of the sulfur surface in the pyrite as well as the typical chalcopyrite grain size formation [39].

Other studies on Au and associated minerals have concentrated on the valence band electronic structure using nanoparticle sizes onto NiO, TiO$_2$, and Al$_2$O$_3$ [24, 27, 36, 37] using ultraviolet photoelectron spectroscopy (~40 eV). Also, conventional Al K$\alpha$ X-ray XPS (1486 eV), was used to study the electronic structure of Au nanoparticles [on carbon [40-44]. The results from these studies indicate that a metal-insulator transition occurs as a function of size between Au atoms which is attributed to quantum size effects.

In order to understand the relationships between Au and associated minerals, there is the need to study chemical bonding and properties of elemental species in these indicator minerals. This will aid in determining features of elements' concentration and diffusion during ore formation and flow mechanism in hydrothermal systems ascribe to Au host minerals. Despite increasing knowledge that indicator minerals can host numerous elements, most geologists in the study area have mainly concentrated on host minerals for Au such as arsenopyrite, pyrite, garnet and quartz that occur in orogenic and sediment hosted disseminated gold deposits known as Carlin type [45].

Bayari *et al.* [46] used X-ray florescence spectroscopy, XRD and inductively coupled plasma mass spectrometry to identify pathfinder elements and their associated pathfinder minerals found in the mineralized regolith profiles at the Bole-Nangoli gold belt in the north-eastern Ghana. Also, Nude *et. al.* [47] used a multivariate statistical approach to identify pathfinder elements for gold from the northwest of Ghana. In both cases the trace elements such as Fe, Mn, Ag, As, Cu, Zn, Ni, Pb, etc. were identified and appeared to be associated with Au and suitable as pathfinder elements of Au in the study areas.

Up till now, there is little or no work on pathfinder elements and minerals in Ghana using XPS and EDX, and in particular, in the Kubi Gold mining area. Rock and mineral identifications leading to Au discovery in this exploration area have been based on geophysical techniques and geochemical analysis (soil/sediments sampling) in laboratories as well as geologists' experiences in mineral identification. However, these procedures are sometimes misleading and may cause delays in projects and high financial loss to mineral exploration activities. Therefore, there is a need to introduce faster and more robust techniques for effective identification of pathfinder elements and minerals in the mining industry.

In this present work, fine-grained sediments (mining waste) from artisanal small-scale mining sites are investigated by surface sensitive XPS and bulk sensitive EDX to quantify the Au content and other important elements that act as indicator minerals of gold. We investigate Au samples from alluvia deposit to identify individual elements in order to gain corresponding insight into surface reaction mechanisms as well as detailed properties of precipitated Au particles observed during the hydrothermal formation of the ore deposit. These identifiable pathfinder elements are then utilized to infer the original indicator host minerals in the area.

## 2. Experimental details

### 2.1. Description of study area

Samples for the study were taken from the alluvia small scale mining site located at Dunkwa-on-Offin district in the central region of Ghana. Figure 1 shows the district that is located within latitudes 5° 30' and 6° 02' north of the equator and longitudes between 1° W and 2° W of the Greenwich Meridian [48]. Close to the proximity of the Kubi Gold Barbados is the alluvia mining site, where the crushed and washed samples were collected for the project. The area is drained by a number of rivers and streams including the Offin River as the main river source which is 90 m above sea level forming a boundary between the Ashanti region and the Central region with the Pra River [49]. The area consisting of volcanic belts, basement rock, and sedimentary basins with much igneous (volcanic) activity and two orogenic events genuinely follows the geology of Ghana, leading Au hosted quartz veins to dip steeply in shear zones with Birimian basins of sulfur rich minerals. The gold at the Kubi mining area is *free milling* (open cast mining) and occurs within a near vertical 1 km by 1 m to 15 m thick shear confined garnet zone. The ore is estimated to be hosted by 30% garnet, 15% sulfide mineralization, and also occurs as coarse gold in late quartz veins.

In addition, the Offin river has alluvia placer Au deposits in gravels and in quartz-pebble conglomerates of the Tarkwaian deposits [48]. Within the alluvia regime are oxides or weathered rocks of the host minerals, which are mostly iron oxides and mineralized quartz [50].

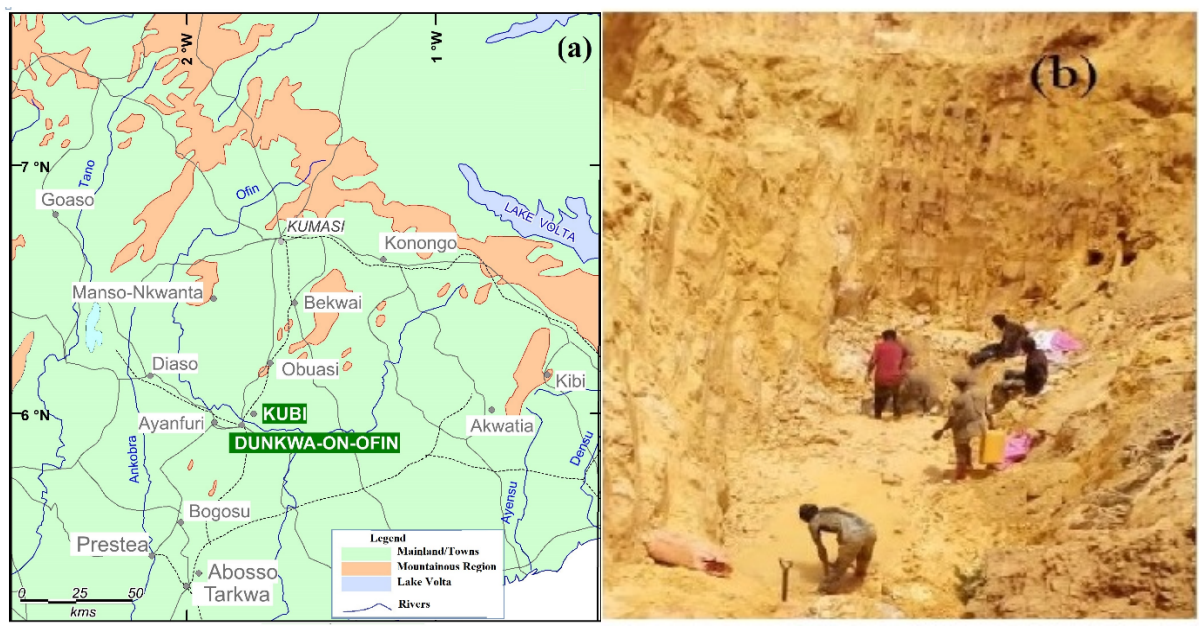

**Figure 1.** (a) Map of Kubi village and surrounding Townships and (b) Picture of sampling area of "Kubi Gold Barbados", Dunkwa-on-Offin. [Map and photo by G.K. Nzulu].

*2.2. Behavior of gold in hydrothermal systems*

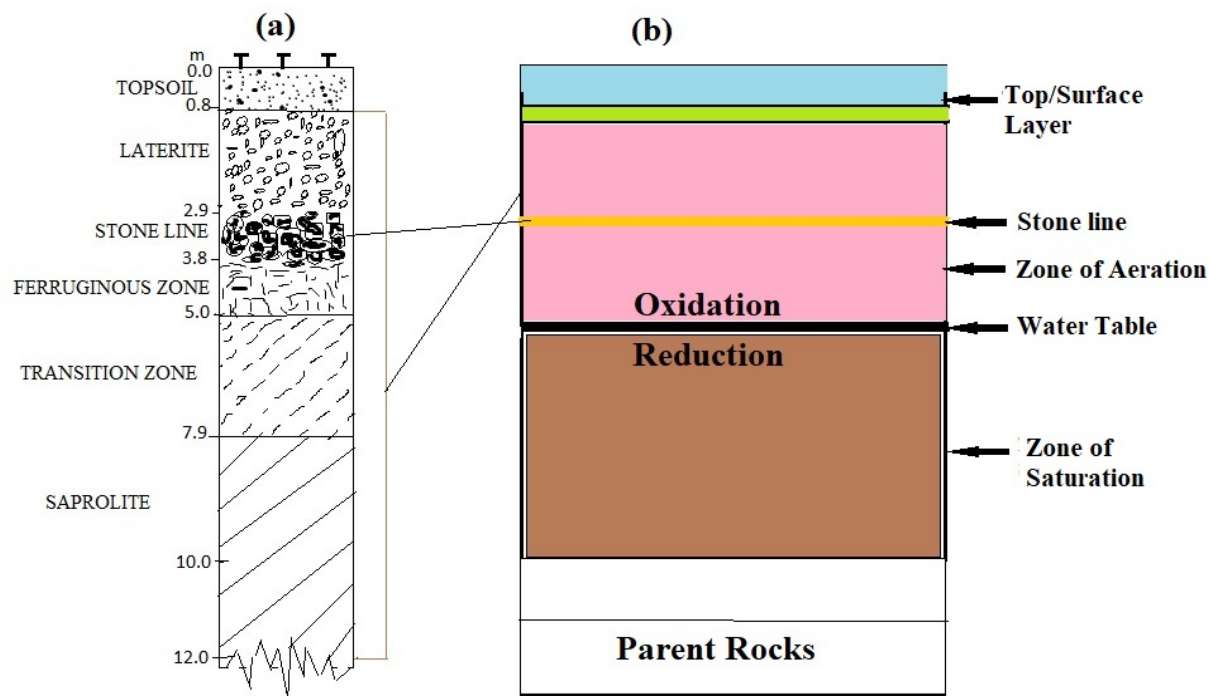

**Figure 2.** (a) Schematic section of the Kubi soil profile (b) Schematic section of ground profile showing Au auriferous lodes.

Figure 2 (a) shows schematic profile of the sampling area (Kubi soils). The saprolite (altered by hematite and limonite/goethite) that dominates the profile grades upwards into the transition and continues to the clay zone, which consist of ferruginous materials, mostly of goethite, kaolinite, and chlorite. Within *the stonel line* horizon are fragments of quartz (pebble and cobbles) and gravels mixed with ferruginous clay that have been formed by weathering profile. The Au in this soil is due to weathering activities and in association with sulfide group of minerals and Fe-oxides. The smaller Au grains found a few meters on the surface occur beyond the *stone line* lateritic zone through chemical dissolution [51]. Thus, beyond the *stone line*, Au mineralization occurs with much coarser Au grains as a result of pedoturbation (soil horizon mixing) that goes downward due to migration of fragments from the stone line through the transition zone and to the saprolite and vice versa.

Figure 2 (b) shows the oxidation and reduction processes that occur when pathfinder minerals undergo chemical dissolution to deposit gold above and below the water table. Au formed on the oxides (near surface) and on external particles, are usually from a mixture of chemical reactions and microbiological activity in groundwater of sedimentary origin of pH ranging from 6 to 9. During flow regime within the hydrothermal system (zone of aeration), the rich sulfide minerals (pyrite group) and the oxide groups can release metastable thiosulphate ions ($S_2O_3^{2-}$) under oxidation process to form composite material with Au under near-neutral-alkaline pH conditions; to trigger a dissolution of Au particles within the sediments/minerals [52, 53]. These thiosulphate ions can decompose after a short period to aid in the oxidation of sulphate ions formed above the water table, or to aid in the reduction of dissolved hydrogen sulfide ions formed below the water table (zone of saturation) [53]. Sulphate minerals found in underground water undergo a reduction process to form bisulphide ions ($HS^-$) near the water table to assemble Au in sedimentary environment under near pH conditions. This same chemical process can facilitate the precipitation of Ag with Au under favorable pH conditions, to deposit gold together with little quantity of Ag [52, 53]. Thus, under alkaline conditions, the most soluble complex of the Au-Ag-S-O system is the $Ag(S_2O_3)_2^{3-}$ ion, such that the separation of Ag and other elemental species from Au at high pH is always a possibility [54].

*2.3. Sample preparation*

A 10 m depth extracted sediment containing Au samples from the artisanal small-scale mining site in Dunkwa-on-Offin has been investigated in this study. The sample was collected by one of the authors (G.K.N) with the aid of a geological hammer into a sample collection bag. The composite "concentrate sample" of weight 1.90 kg as shown in Figure 3a, was divided into two parts, where one portion (1.20 kg) was refined into pure solid Au (bulk Au), whilst the other powder sample (0.70 kg) was subjected to panning i.e., washing (Figure 3c) and extraction of residual sediment of $Fe_3O_4$, $Fe_2O_3$, and other minerals to obtain an untreated Au powder concentrate of weight of 80 g (Figure 3d). This untreated powder sample was later separated into coarse-grained (0.85 g) and fine-grained (1.20 g) with the aid of a tweezer and water panning (via gravity) respectively (the residual 77.95g containing sand and gravel was disposed). The impure powder sample (3d) was first investigated by XPS after which a tweezer was used to pick the average grained size particles as coarse-grained powder sample. The final residual sample was water-panned for tiny particles of Au, quartz, and pyrite group as fined-grained powder samples.

The final three samples; namely solid Au (bulk), untreated Au concentrate powder, fine and coarse-grained Au concentrates in figures 3b, 3e, and 3f, respectively, in addition to a standard thin film Au sample in Figure 3g, were examined by XPS to reveal distinct features of the materials. Particle/grain sizes of the powder samples ranges from 0.05 cm to about 0.2 cm in maximum dimensions of which most were hoppered single crystals. The nature of materials indicates that most Au from this artisanal area come from weathered $SiO_2$ veins hosted by coarse grained igneous and meta-volcanic rocks with mineralization controlled by garnetiferous horizon of fine-grained Au associated with sulfide minerals and occasional carbonates [50].

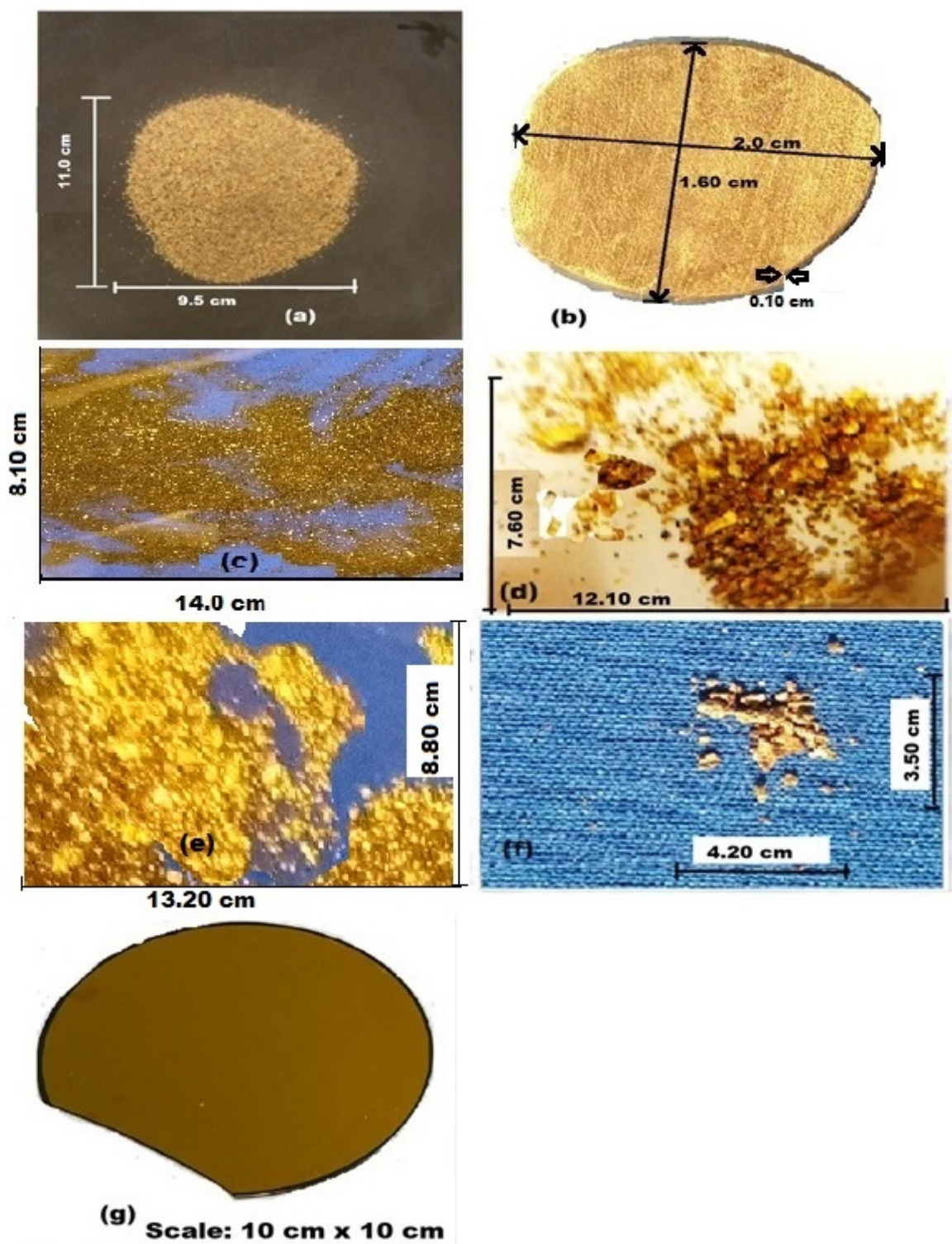

**Figure 3.** (a) Dry residual sample (tailing) from an artisanal mine site, at Dunkwa-On-Offin (b) Refined part of sample into pure Au nugget of one gram and 22 carats (measured with a digital electronic Au purity Analyzer D H 300 K from VTSYIQ). (c) Panned sample of Au with other impurities (d) untreated ore powder (e) fine-grained concentrate Au, (f) coarse-grained concentrate powder and (g) Au film on wafer substrate. [Photo by G.K. Nzulu].

*2.4. XPS measurements*

XPS was used to analyze the elemental composition and chemistry of the powder samples (untreated and fine and coarse-grained Au ore) and reference materials (bulk nugget and a 300 nm thick Au film with a 2.5 nm Ti buffer layer on a Si wafer substrate). The analyses were performed in an Axis Ultra DLD instrument from Kratos Analytical (UK) employing monochromatic Al K$\alpha$ radiation (h$\nu$ = 1486.6 eV) and operated at a base pressure lower than $1.1 \times 10^{-9}$ Torr ($1.5 \times 10^{-7}$ Pa) during spectra acquisition. XPS depth profiles were acquired by sputter-etching with 0.5-keV $Ar^+$ ions incident at an angle of 70° relative to the sample surface normal. The low $Ar^+$ energy and shallow incidence angle were chosen to minimize the effect of sputtering damage in the spectra. The sample areas analyzed by XPS were $0.3 \times 0.7$ mm$^2$ and located in the center of $3 \times 3$ mm$^2$ ion-etched regions. The binding energy (BE) scale was calibrated using the ISO-certified procedure, with the spectra referenced to the Fermi edge in order to circumvent discrepancies associated with employing the C 1s peak from adventitious carbon [55]. Prior to the XPS core and valence band measurements, the structural properties were investigated [56]. From the XRD analysis [56], Au, quartz ($SiO_2$), magnetite ($Fe_3O_4$), and hematite ($Fe_2O_3$) were identified in these samples.

*2.5. EDX measurements*

The bulk, untreated, fine- and coarse-grained concentrate samples were analyzed using SEM/EDX operated at 20 kV. The EDX used in this study is of the Zeiss Supra 35 VP field emission SEM type whose field emission scanning electron microscopy (FESEM) is equipped with an energy dispersive X-ray spectroscopy (EDX) detector, which was used for elemental mapping of the samples to understand the elemental distribution. Samples were prepared by mounting the macrocrystalline samples into aluminum stubs with carbon sticky tabs.

**3. Results and discussion**

*3.1. Core-level XPS*

Figure 4 shows XPS survey spectra of Au in the form of thin film, bulk, and powder (untreated, fine-grained, and coarse-grained) concentrate samples, that are vertically offset for clarity. The observed peaks of the thin film are identified as Au *4f*$_{7/2,5/2}$ at 84 and 87 eV, Au *4d*$_{5/2,3/2}$ at 335 and 353 eV, Au *4p*$_{3/2,1/2}$ at 546 eV and 642 eV, and Au *4s* at 762 eV. The impure powder sample was first investigated by XPS after which a tweezer was used to pick the average grained size particles as coarse-grained powder sample. The final residual sample was water-panned for tiny particles of Au, quartz, and pyrite group as fined-grained powder samples.

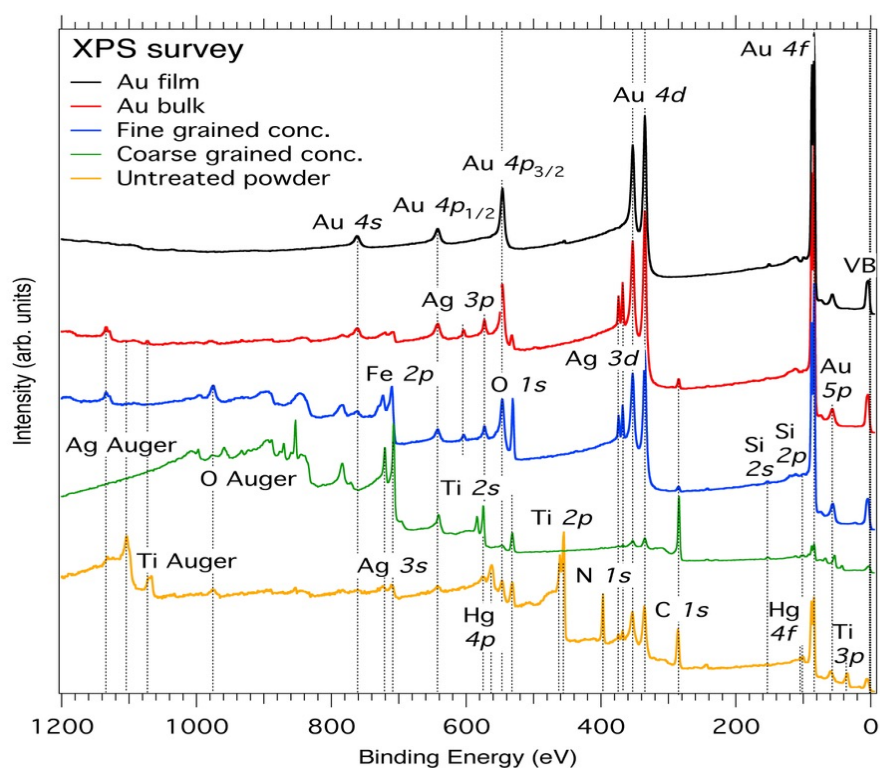

**Figure 4.** XPS survey spectra of film, bulk, and powder (untreated and fine and coarse-grained concentrate) Au. The broken line connects the band edges as well as the apparent spin-orbit-splitting maxima. VB denotes valence band.

In the Au bulk sample, we can also identify all the Au peaks as well as peaks arising from Ag $3d_{5/2,3/2}$ at 368 and 374 eV, Ag $3p_{3/2,1/2}$ at 573 and 604 eV, and Ag $3s$ at 719 eV. In addition, there are C $1s$ and O $1s$ peaks at 285 and 532 eV, respectively. In the untreated concentrate powder sample, we find in addition to the Au peaks, a Ti $3p$ peak at 37.50 eV, Ti $2p_{3/2,1/2}$ peaks at 458 and 461 eV, respectively, and N $1s$ peak at 397 eV as listed in Table 1 and are consistent with reference data [57].

In the present experiment sulfur is very diluted in the samples due to the preparation method that make the detection of sulfur by XPS very challenging (low sensitivity). Only a very weak S $2p$ peak around 165 eV was observed in untreated powder while in the fine-grained sample it was hardly discernable. In the other samples it was below the signal-to-noise level. The S $2s$ cross-section 231 eV is even lower than for the S $2p$ signal.

The Fe in the powder (untreated, fine, and coarse-grained concentrates) and Au bulk samples has $2p_{3/2,1/2}$ peaks located at 706.90 and 720 eV, respectively, consistent with literature [58, 59] and in agreement with reference data [57]. The Fe $2p_{1/2}$ peaks are closely located at the same BE of 719 eV as the Ag $3s$ peak [57].

The measured XPS data of Au $4p_{3/2}$, O $1s$, and C $1s$ in Figure 4 provide complementary information on the cleanliness of the sample surfaces and the state of oxidation and carboniferous (carbonic) contamination [60] as these alluvia samples may contain mine discharge, dissolution of carbonate minerals of well-buffered pH, different calcite saturation indices, and dissolved in oxygen [61].

Figure 5 shows the core-level Au $4f$ XPS spectra of the Au film, Au bulk, and untreated concentrate powder of Au. For the Au film and bulk sample, the $4f_{5/2}$ peaks are located at 87.63 eV and the $4f_{7/2}$ peaks are located at 83.95 eV with a 3.6 eV spin-orbit splitting. These binding energies are similar to the literature values of 87.60 and 84.0 eV [57] for pure Au.

For comparison, in the Au powder sample, the $4f_{5/2}$ is located at 87.77 eV and the $4f_{7/2}$ peak at 84.09 eV. The Au $4f$ peaks are asymmetric, while the shape of the untreated concentrate Au powder and coarse-grained concentrate Au powder spectra show broadening with tails toward higher BE. The chemical shift of 0.14 eV towards higher BE for the powder samples suggests that there is a charge transfer from Au towards the other elements. The high-energy shift can be attributed to the screening effect in comparison to the pure bulk Au and suggests total charge-transfer from Au towards the certain elements.

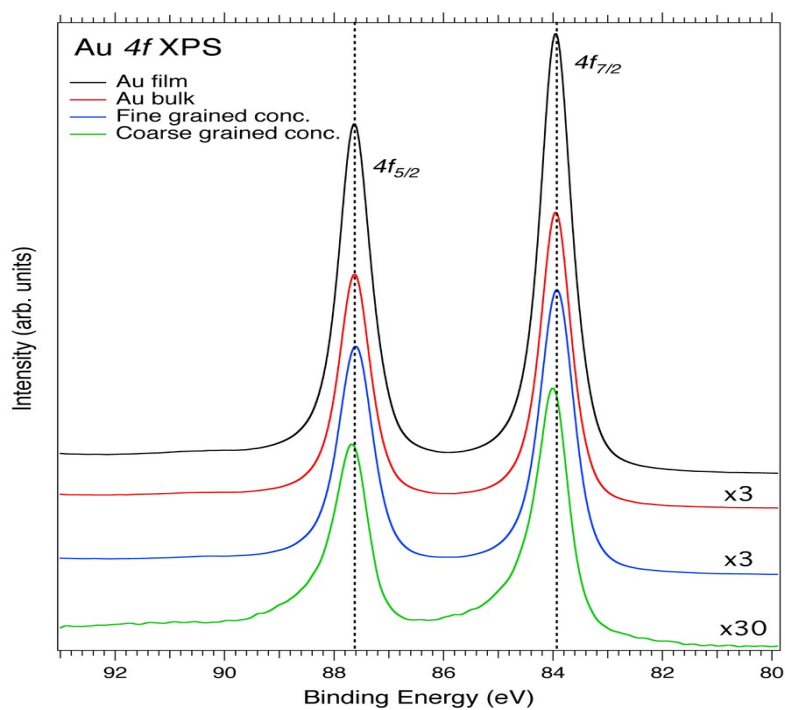

**Figure 5.** Au $4f$ core-level XPS spectra of the Au film, Au bulk, fine-grained Au concentrate, and coarse-grained Au concentrate. The dashed line connects the $4f$ peaks as well as the apparent spin-orbit-splitting maxima. The spectra containing lower Au concentrations were scaled-up by factors of 3 and 30 for comparison.

From the spectra in Fig. 5, the Au 4*f* region has well separated spin-orbit components of 3.6 eV which is close to Au metal (Δ=3.7eV) and is in agreement with literature [62]. The Au 4$f_{7/2}$ peak at 84.0eV serves as a useful BE reference for Au metal [57] and expected to shift (BE shift) for smaller cluster samples to Au nanoparticle sizes.

The spectra from Figure 6 shows the Ag $3d_{5/2,3/2}$ core level XPS of the bulk and untreated Au powder, fine-grained Au, and coarse-grained Au concentrate samples. The peaks for the Ag $3d_{3/2,5/2}$ levels are located at 374.0 eV and 368.0 eV respectively, with respect to the Fermi level ($E_F$) in reasonable agreement with literature data [57].

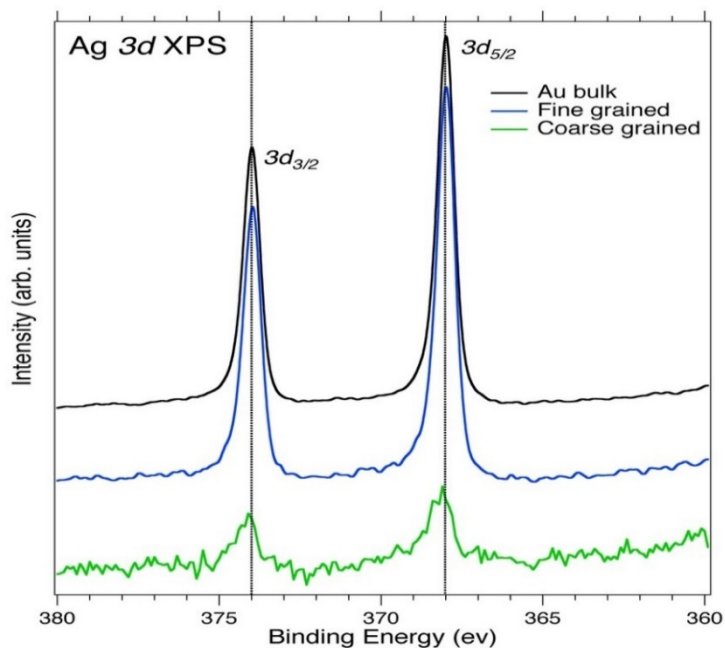

**Figure 6.** Ag $3d_{5/2,3/2}$ core level XPS spectra of the bulk Au and powder (unprocessed and processed) Au samples. The dashed line connects the peaks as well as the apparent spin-orbit-splitting maxima.

The Ag $3d_{5/2,3/2}$ spectra in Figure 6 show well-separated spin-orbit components ($\Delta_{metal}$ = 6.0 eV) and the peaks display asymmetric peak shape for Ag metal and also show loss features to higher BE side of each spin-orbit component of Ag metal [63]. The bulk Au and fine-grained Au concentrate powder, show higher intense peaks due to the high number of Ag atoms present in the samples, having interaction with unfilled electron levels of the valence band above $E_F$. The lower intense peaks and the loss of features for the coarse-grained Au concentrate sample suggest negligible number of Ag atoms present in the material. The smaller BE shifts observed in the spectra suggest the presence of oxides in the materials causing the Ag $3d_{5/2,3/2}$ peaks to broaden with respect to metal peaks as a results of electronegativity difference (lattice potential, work function changes, and extra-atomic relaxation energy) which causes a charge transfer from the metal atoms to the ligands [64, 65, 66].

The reason that Au and Ag alloy to form "*electrum*" in the bulk and powder samples is due to their partly filled d bands (d-hole count effect) and having the same fcc structure makes it possible to exhibit relativistic effects where both bands display correct spin-orbit-splitting signature. The structural changes that occur during the formation of the electrum affect the d-hole count in both Ag and Au and increases with decreasing Au concentration which gives a solid correlation with the ostensible Au *5d* spin–orbit splitting, lower d-band energy and the Au 4$f_{7/2}$ BE shift [67].

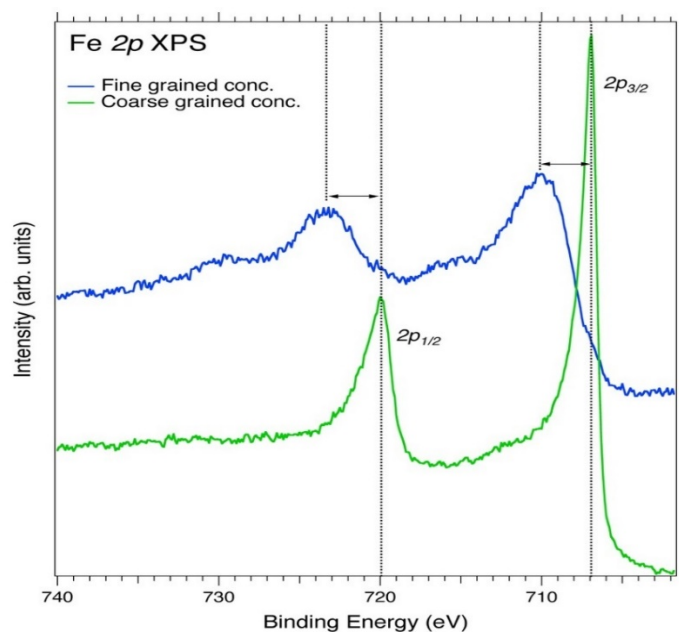

**Figure 7.** Fe *2p* core-level XPS spectra of the fine-grained Au concentrate and coarse-grained Au concentrate samples. The broken line connects the peaks as well as the apparent spin-orbit splitting maxima.

From the XPS data the Fe *2p* in the sample corresponds to pure Fe metal with the *2p$_{3/2}$* peak at 710.9 eV and the spin-orbit splitting of 13.6 eV. These values are close to that obtained from coarse-grained Au concentrate sample values with *2p$_{3/2}$* = 706.9 eV and spin-orbit components ($\Delta_{metal}$=13.1eV). The Fe *2p* peaks in Figure 7 show asymmetric shape for metal, but the Fe 2p$_{3/2}$ spectrum is not well resolved to exhibit multiplet splitting. A high BE shift due to screening in comparison to pure metal show that there is charge transfer from the metal atoms to ligands. If the Fe in the samples originated from the oxide zone (FeO and $Fe_2O_3$, and $Fe_3O_4$), the peaks would have significantly shifted towards higher BE, but the actual shift towards lower BE (~710.9 eV) suggest, the peaks originated from pure Fe metal or pyrite ($FeS_2$) as opined by Biesinger *et al*. [68]. Iron compounds may possess high or low spin, with Fe (III) compounds exhibiting high-spin which lead to complex multiplet-split [68]. Fe (II) compounds, however, have both high or low spin with spectra from low spin lacking multiplet splitting and attributed to marcasite ($FeS_2$) or pyrite group of compounds [68]. According to Uhlig *et al*., (2001), for air-exposed $FeS_2$, the $Fe^{2+}$ is in a low-spin configuration and the broad structure towards higher BE (710.5 eV) of the main Fe *2p$_{3/2}$* peak, is due to oxidation of $Fe^{3+}$ states of the surface [69]. For smaller marcasite peak, the higher BE (708.4 eV) is due to electronic-deficient of $Fe^{2+}$ sites, formed by breaking of Fe-S bonds [69]. This higher BE (708.4 eV) and lower for marcasite $FeS_2$ is in closed agreement with the peak at (706.9 eV) from the XPS spectra in Figure 7, which is consistent with reference data [70] for $FeS_2$. It can be deduced that the deficient main Fe *2p$_{3/2}$* peak in the samples from Kubi Gold project contain pure metal Fe and $FeS_2$, since $FeS_2$ is one of the main indicator minerals in the mining area [56]. As telluric iron is extremely rare, we anticipate that the pure metallic Fe observed by our XPS measurements originates from iron oxides are readily reduced in the argon monomer sputter cleaning process.

The assigned peak values of Ti in Table I are in good agreement with reference data [62] and consistent with literature [71-74]. These abundant titanium in the powder sample is believed to be contained in fine-grained aggregates of hematite [75, 77] which is a gangue mineral in this mining area [56]. The silicate garnet minerals which control the Au mineralization in Kubi contains some percentage of $TiO_2$ [59]; whilst other occasionally encountered silicate minerals such as biotite, hornblende, and the abundant silica (quartz), as well as some oxide group in this area containing little $TiO_2$ are responsible for the significant Ti in the untreated Au sample. The assigned peak values of Si in Table I are in close agreement with reference data [62] and consistent with literature [78, 79] and is believed to be contained in the $SiO_2$ which is also an indicator mineral in the mining area. Finally, the assigned peak positions of Hg, a trace element in the untreated powder sample which occur at binding energies of 104 eV and 577 eV, overlap with Si 2p and Ag 3p peaks. These values are in agreement with reference data [62] and consistent with literature [80, 81].

As shown in Table II, quantitative XPS analysis shows that the Ag content in the Au bulk nugget is about 6at.% whilst the C, O, and Fe contents are 3at.%, 2at.% and 1at.% respectively. In the fine-grained Au concentrate powder sample, the Ag content is at.% half of that of bulk sample, while the level of C and O is significantly higher.

**Table I.** Elements identified in the analysis of XPS core levels for Au bulk and powder (untreated, fine-grained and coarse-grained) Au concentrate samples.

| Element | Spectral line | Formula | Energy (eV) | Parent Mineral |
|---|---|---|---|---|
| Au | 4s | Au | 762.0 | Gold |
| Au | $4p_{3/2,1/2}$ | Au | 546.0/642.0 | Gold |
| Au | 4d | Au | 335.0 | Gold |
| Au | $5p_{3/2}$ | Au | 57.20 | Gold |
| Au | 4*f* | Au | 84.0/87.0 | Gold |
| Ti | 3p | $TiO_2$ | 37.50 | Hematite, garnet and other silicate minerals |
| Ti | 2s | $TiO_2$ | 561.00 | Hematite, garnet and other silicate minerals |
| Ti | $2p_{3/2, 1/2}$ | $TiO_2$ | 458.0/464.19 | Hematite, garnet and other silicate minerals |
| Si | 2p | $SiO_2$ | 102 | Quartz |
| Si | 2s | $SiO_2$ | 153.0 | Quartz |
| S | 2p | S | 163 | Au-S/$FeS_2$ |
| S | 2s | S | 231 | Au-S/$FeS_2$ |
| Ag | $3p_{1/2, 3/2}$ | Ag | 604.0 - 573.0 | Silver |
| Ag | $3d_{3/2, 5/2}$ | Ag | 368.0 - 374.0 | Silver |
| Ag | 3s | Ag | 719.0 | Silver |
| N | 1s | N | 397.0 | Nitrogen |
| O | 1s | O | 532.0 | Oxides |
| C | 1s | C | 285.0 | Carbon/graphite |
| Fe | $2p_{3/2}$ | Fe/Cu | 706.90 | Iron/$FeS_2$ |
| Fe | $2p_{1/2}$ | Fe/Cu | 720.0 | Iron/ $FeS_2$ |
| Hg | 2p | Hg | 104 | Hg |
| Hg | $4p_{3/2}$ | Hg | 577 | Hg |

**Table II.** Results of quantitative analysis of XPS core levels for bulk Au, fine-grained and coarse-grained Au concentrate samples.

| System | Au4*f* (at.%) | Ag3d (at.%) | C1s (at.%) | O1s (at.%) | Fe2p (at.%) | Mn (at.%) |
|---|---|---|---|---|---|---|
| Au bulk | 88 | 6 | 3 | 2 | 1 | - |
| Fine-grained Au | 32 | 4 | 7 | 32 | 24 | 3 |
| Coarse-grained Au | 6 | 1 | 27 | 34 | 26 | 6 |

*3.2. Chemical bonding*

Figure 8 shows valence band XPS spectra of an Au film, Au bulk, unprocessed ore powder and processed (coarse and fine grained) ore powder of Au in comparison to calculated Au spectra. The agreement with the calculated peak positions in the valence band is improved when experimental lattice parameters (*a* = 4.078 Å) are used while the theoretical (relaxed) lattice parameter of 4.156 Å is larger and results in peak positions that have too low BE.

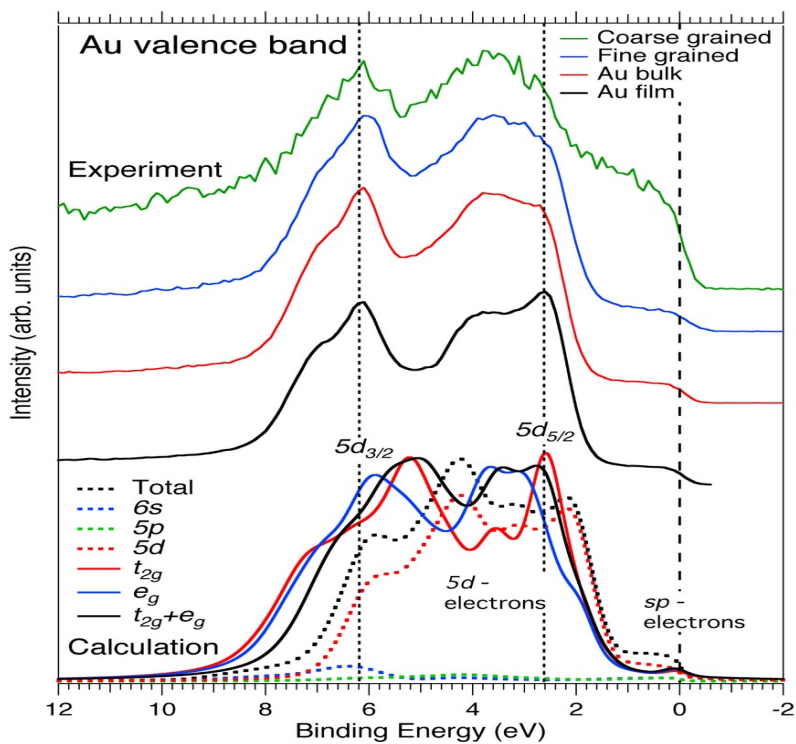

**Figure 8.** Experimental and calculated valence band spectra of film, bulk, fine-grained, and coarse-grained powder. The broken line connects the *d*-band edges and the apparent spin-orbit splitting maxima.

The valence band spectra in Figure 8 show the coordination complex Au (transition metal) with an octahedral geometry where the *d* orbitals split into $t_{2g}$ set (lower energy) and an $e_g$ set (higher energy) [82, 83]. From the spectra, the $e_g$ set moves towards higher BE due to the highly involvement of the d orbitals in the $e_g$ set in metal-ligand sigma (σ) interaction. These orbitals in the $e_g$ are directed towards the axis where they are opposed by strong repulsive forces from ligands and hence, they acquire higher energy than the $t_{2g}$; thus, favoring covalent bonding in $e_g$ towards higher energy [83]. The shift in BE along the $e_g$ set for the powder sample (coarse-grained) is an indication of covalent bonding of Au with the oxides (ligands) in the powder samples and these extend beyond the $e_g$ to higher BE. Au has a face centered cubic (fcc) crystal structure, and its bonding is a mixture of covalent, ionic, and metallic bonds. Due to symmetry considerations, the covalent contribution consists of Au $e_g$-ligand 2p (*pd*-σ), Au $t_{2g}$-ligand 2p (*pd*-π), and Au-Au $t_{2g}$ (*dd*-sigma) bond regions that can be observed in the valence band.

The lower energy $t_{2g}$ orbitals do not favor sigma interactions with ligands. The $t_{2g}$ symmetry favors the participation of π interactions with ligands due to their weaker interactions in metal complexes [83]. This means the $t_{2g}$ orbitals are completely metal based and contribute to metallic bonding. From the valence band spectra, the pronounced peaks at $t_{2g}$ for Au film, bulk, and fine-grain powder indicate metallic bonding with other metals, whilst the coarse-grained powder show weaker peaks for metallic bonding due to the high level of oxidizing materials.

The spectra result in Figure 8 for the Au film, bulk, fine, and coarse-grained indicate that the BEq of the Au $5d_{3/2}$ component of the valence band remain almost unaffected, whilst the Au $5d_{5/2}$ shifted away from the $E_F$ for decreased Au sample size, the top of the d-band is observed to move away from the $E_F$. Thus, the Au $5d_{5/2}$ of fine and coarse-grained samples due to the small grain size shifted away from the $E_F$ as well as the top of the d-band [84-86].

The spectra show that for smaller Au powder samples (fine and coarse-grained concentrates), there is an increase in the BE side on the asymmetry component while the Auger yield tend to decrease [55]. The spectra of the powder samples show disturbance peaks towards the Auger yield that are attributed to the state of oxidation the samples are subjected to. This shift towards higher BE of the Au $4f_{7/2}$ core-level for smaller cluster size (powder samples) relates to the valence band narrowing, ascribes a general shift, and interpreted to be dependance on the initial electronic state structure and the final relaxation state processes as a result of photoemission [85,87,88].

For the Au (film, bulk, fine and coarse-grained) samples, the valence band width and the binding energies of the two Au $5d$ and $4f_{7/2}$ peaks depend solely on the mean coordination number [83-85]. The $4f$ photoemission from Au split into two distinct peaks namely, Au $4f_{7/2}$ and Au $4f_{5/2}$ with different BE as a result of spin-orbit coupling effects corresponding to final states with angular momentum j+ = ⌊ + 1/2 = 7/2 and j- = ⌊ + 1/2 = 5/2 respectively. It is seen from the Au $4f$ spectra that the Au $4f_{7/2}$-to-$4f_{5/2}$ peak ratios are not the same but varies between the four different samples. The useful information one can obtain from the study of the BE shift of Au $4f_{7/2}$, full width at half maximum (FWHM) of the Au $4f_{7/2}$ peak and valance band width is the spectra response to average sample size, in terms of peak broadening and shift towards higher BE, as seen in the Au bulk and powder samples [91-93].

*3.3. EDX results*

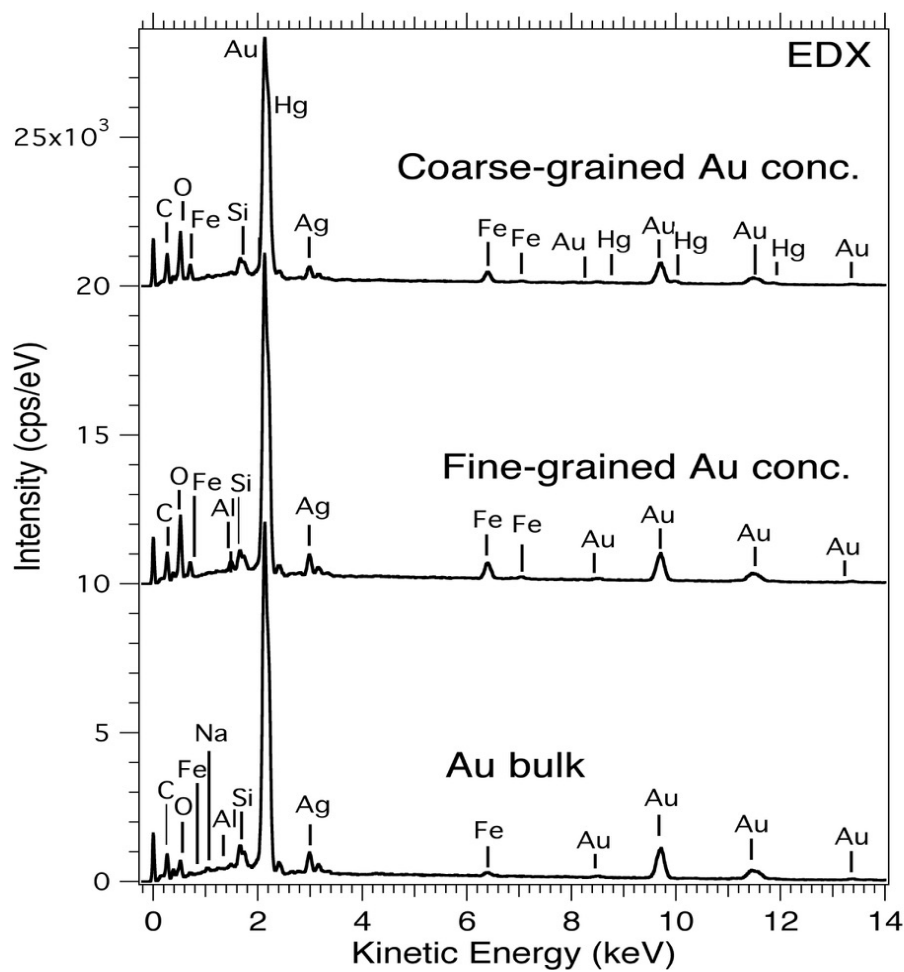

**Figure 9.** EDX spectra of bulk Au, fine-grained and coarse-grained Au concentrate powder samples.

Examination under the EDX of the Au powder samples (fine and coarse-grained) show that more than 50at. % of the Au is present in the grains and these are associated with the pathfinder minerals garnet, hematite, and quartz. The EDX results in Figure 9 and Table III also reveal other elements such as Al and Na whose peaks are not easily identified by the surface sensitive XPS measurements since light elements (N, O, C, and surface oxides including $Fe_2O_3$, $Fe_3O_4$, and $SiO_2$) are removed by differential sputtering of $Ar^+$ ions, whilst bulk sensitive EDX does not require any surface cleaning.

Table III. Elements identified in the analysis of EDX for bulk Au, powder (fine-grained and coarse-grained) Au concentrate samples.

| Sample | Bulk Au (at%) | Powder fine-grained (at.%) | Powder coarse-grained (at.%) |
|---|---|---|---|
| Element | | | |
| C | 11.7 | 9.50 | 23.3 |
| O | 7.44 | 14.4 | 22.1 |
| Si | 1.17 | 0.58 | 2.55 |
| Fe | 1.82 | 4.39 | 3.40 |
| Ag | 6.39 | 6.40 | 4.58 |
| Au | 71.2 | 64.2 | 42.6 |
| Al | 0.38 | 0.57 | - |
| Hg | - | - | 1.55 |

The Fe in these alluvia Au samples is believed to be contributions from the presence of pyrite, pyrrhotite, chalcopyrite and the "gangue mineral" magnetite [56]. According to Moslemia et al. [94] and Corkhill et al. [95] oxidation of minerals such as $Fe_3O_4$ and $CuFeS_2$ (pyrite group of minerals) under conditions such as pressure, temperature, pH, particles size, concentration and electrochemical mechanisms result in the release of ferrous ions and other constituents [94, 95]. Hence, it can be inferred that, the Fe in these samples is from pure elemental Fe and $FeS_2$. In our analysis we have prepared the samples (fine and coarse grained) from the untreated powder so that almost all sulfur containing grains were discarded. This is due to the fact that the sulfide minerals get washed away during panning and washing since they are much lighter than the Fe and Au metals that settle at the bottom of the pan.

In contrast to previous observations by XRD [56], where the crystal structures were identified, we find that there are significant changes in the chemical bonds in the investigated materials. The XPS analysis confirms that there is covalent bonding between the Au atoms and ligands in particular for coarse-grained, while the Au bulk and fine-grain powder indicate metallic bonding with other metals as shown in the valence band spectra in Fig. 8. The ratio between metallic and covalent bonding depends on the level of oxidizing materials presence in the vicinity of the Au atoms. These oxidizing materials from the near surface (oxide zones) consist of the gangue minerals $Fe_2O_3$, $Fe_3O_4$, $SiO_2$, garnet, and other silicate group of minerals that under favorable conditions, undergo oxidation and reduction processes to release metastable thiosulphate ions ($S_2O_3^{2-}$) and bisulphide ions ($HS^-$) to trigger dissolution of Au, Ag, and other metals within the hydrothermal systems [52, 54].

These identified elemental species Si, Ag, Fe, Al, N, O, Hg, and Ti can act as pathfinder elements in the host minerals of the Kubi Au ore and can lead to exploration success in this mining area. Despite similarities in geology, mineralogy, and structural setting of Kubi Gold project with that of Northern belts of Ghana, there exist differences in pathfinder elements due to weathering, erosion, chemical, and landscape processes leading to differences in regolith (loose materials covering solid rock material) [46, 47]. In the Kubi Gold area, the association of Ti and Fe with Au is unique and differs from what exists in the northern Ghana. The saprolitic and ferruginous clay that have been formed by weathering profile are dominated by garnet (that contains $TiO_2$), Fe-oxides, and iron sulfide minerals through secondary process and capable of hosting Au mineralization.

The most interesting part of this study is the scope of economic benefits to be considered from the 1.90 kg artisanal small-scale sediments (soil samples or mining wastes), which gave 1.0 g of Au bulk nugget (22 carats measured with a digital electronic Au purity Analyzer DH 300 K from VTSYIQ), 0.85 g of coarse-grained, and 1.20 g of fine-grained concentrate samples. These samples are found to contain high percentage of Au as well as metallic and semi-metallic elements such as Ag, Fe, Al, and Ti that are important mineral commodities of economic values.

These new findings from the near surface sediments call for further investigations into the ore-body of deep drilled holes where materials have been subjected to high temperatures and pressures. We anticipate that there will be depth dependent composition with much deformations and defects due to hydrothermal and physio-chemical processes amongst others in the ore-body.

## 4. Conclusions

By combining core-level and valence band X-ray photoelectron spectroscopy with EDX electronic structure calculations, we have investigated the electronic structure and chemical bonding in various powder Au samples that remain after Au panning in comparison to bulk Au metal and Au film. For the impure Au powder (fine-grained and coarse-grained concentrate) samples, there is a chemical shift of 0.14 eV for the Au $4f_{7/2}$ and $4f_{5/2}$ peak positions towards higher energies compared to the metal values of 83.95 and 87.60 eV due to screening effect and charge transfer, indicating bonding between Au atoms and the surrounding ligands in the indicator minerals. This study demonstrates that indicator and pathfinder methods have a broader application which includes the Kubi Gold project, a fact that is not commonly known for artisanal small-scale mining wastes (alluvia small-scale mining sampling). The indicator minerals in this alluvia small-scale mining are part of a large group of indicator minerals that can be used to explore a broad range of Au deposits and other commodities in this mining area.

From the study, we identified Si, Ag, Fe, Al, N, O, and Ti that act as pathfinder elements and can be inferred to indicator minerals such as quartz, hematite, pyrite (marcasite), garnet, and other occasional silicate minerals such as biotite and hornblende. The results also demonstrate that the mobility of Au and pathfinder elements are very high within the oxide zones (near surface) and these elements bond to the indicator minerals hosting the Au in the mining site.

Quantitatively, we find that, the fine-grained Au concentrate has relatively high amount of pure Au that form electrum with significant amount of Ag and this finding makes mining waste an economic viable one to consider. The significant percentage of oxygen in the powder samples are due to contaminations from the fresh alluvia samples, whilst the C is attributed to the carbonic or graphitic alterations in the Kubi mining site.

Finally, we conclude that, fine-grained sediments considered to be "mining waste", could be of economic importance to miners and investors and an important material to exploration or mining geologist to apply XPS, EDX, geochemistry analysis, and other techniques, to elucidate the indicator elements associated with Au host minerals.

**Acknowledgments:** We acknowledge support from the Swedish Government Strategic Research Area in Materials Science on Advanced Functional Materials at Linköping University (Faculty Grant SFO-Mat-LiU No. 2009 00971). The computations were enabled by resources provided by the Swedish National Infrastructure for Computing (SNIC) at the National Supercomputer Centre (NSC) partially funded by the Swedish Research Council through Grant Agreement No. 2016-07213. M.M. also acknowledges financial support from the Swedish Energy Research (Grant No. 43606-1) and the Carl Tryggers Foundation (CTS20:272). Asante Gold Corporation is acknowledged for funding G. K. N.'s industrial PhD studies at Linköping University, Sweden.

**References**

[1] Smyth, J. "Mineral Structure Data". Garnet. University of Colorado, 2007.
[2] Klein, C.; Hurlbut, C. S. Jr.. Manual of mineralogy: (after James D. Dana) (21st ed.). New York: Wiley. 1993, pp. 451–454.
[3] Reducing mercury use in artisanal and small-scale gold mining, A practical guide, UNEP, 2012.
[4] McClenaghan M. B.; Parkhill M. A.; Pronk A. G.; Seaman A. A.; McCurdy M. W.; & Laybourne M. I. Indicator mineral and geochemical signatures associated with the Sisson W–Mo deposit, New Brunswick, Canada, Geochemistry: Exploration, Environment, Analysis, 2017, 17, 297–313.
[5] Noble, R.R.P.; Gray, D.J. & Gill, A.J. Field guide for mineral exploration using hydrogeochemical analysis, CSIRO Earth Science and Resource Engineering, 2011.
[6] Tauson V. L.; Pastushkova T. M. and Bessarabova O. I. On limit concentration and manner of incorporation of gold in hydrothermal pyrite. Russ. Geol. Geophys, 1998, 39, 932–940.

[7] Tauson V. L. Gold solubility in the common gold-bearing minerals: experimental evaluation and application to pyrite. Eur. J. Miner, 1999, 11, 937–947.

[8] Mycroft J. R.; Bancroft G. M.; M, cIntyre N. S. and Lorimer J. W. Spontaneous deposition of gold on pyrite from solution containing Au (III) and Au(I) chlorides: part I, a surface study. Geochim. Cosmochim. Acta 1995, 59, 3351–3365.

[9] Tauson V. L., Mironov A. G.; Smagunov N. V.; Bugaeva N. G. and Akimov V. V. Gold in sulfides: state of the art of occurrence and horizons of experimental studies. Russ. Geol. Geophys, 1996, 37, 1–11.

[10] Simon G.; Huang H.; Penner-H. J. E.; Kesler S. E. and Kao L. S. Oxidation state of gold and arsenic in gold-bearing arsenian pyrite. Am. Mineral, 1999a, 84, 1071–1079.

[11] Simon G.; Kesler S. E. and Chryssoulis S. L. Geochemistry and textures of gold-bearing arsenian pyrite, Twin Creeks Carlin type gold deposit, Nevada. Implications for gold deposition. Econ. Geol, 1996b, 94, 405–422.

[12] Scaini M. J.; Bancroft G. M. and Knipe S. W. An XPS, AES, and SEM study of the interactions of gold and silver chloride species with PbS and FeS2: comparison to natural samples. Geochim. Cosmochim. Acta, 1997, 61, 1223–1231.

[13] Widler A. M. and Seward T. M. The adsorption of gold (I) hydrosulphide complexes by iron sulphide surfaces. Geochim. Cosmochim. Acta, 2002, 66, 383–402.

[14] Cook N. J. and Chryssoulis S. L. Concentrations of ''invisible gold'' in the common sulfides. Can. Mineral. 1990, 28, 1–16.

[15] Fleet M. E. and Mumin A. H. Gold-bearing arsenian pyrite and marcasite and arsenopyrite from Carlin Trend gold deposits and laboratory synthesis. Am. Mineral, 1997, 82, 182–193.

[16] Friedl J.; Wagner F. E. and Wang N. On the chemical state of combined gold in sulfidic ores: conclusions from Mo¨ssbauer source experiments. Neues Jahrb Mineral, -Abh, 1995, 169, 279–290.

[17] Genkin A. D.; Bortnikov N. S.; Cabri L. J.; Wagner F. E.; Stanley C. J.; Safonov Y. G.; McMahon G.; Friedl J.; Kerzin A. L. and Gamyanin G. N. A multidisciplinary study of invisible gold in arsenopyrite from four mesothermal gold deposits in Siberia, Russian Federation. Econ. Geol, 1998, 93, 463–487.

[18] Cabri L. J.; Newville M.; Gordon R. A.; Crozier E. D.; Sutton S. R.; McMahon G. and Jiang D.-T. Chemical speciation of gold in arsenopyrite. Can. Mineral, 2000, 38, 1265–1281.

[19] Velasquez P.; Leinen D.; Pascual J.; Ramos-Barrado J. R.; Grez P.; Gomez H.; Schrebler R.; Del Rio R. and Cordova R. A Chemical, Morphological, and Electrochemical (XPS, SEM/EDX, CV, and EIS) Analysis of Electrochemically Modified Electrode Surfaces of Natural Chalcopyrite (CuFeS2) and Pyrite (FeS2) in Alkaline Solutions, J. Phys. Chem. B, 2005, 109, 4977-4988.

[20] Harmer S. L.; Thomas J. E.; Fornasiero D. and Gerson A. R. The evolution of surface layers formed during chalcopyrite leaching, Geochemical et Cosmochimica Acta 2006, 70, 4392–4402.

[21] Harmer, S. L., Pratt, A. R., Nesbitt, W. H. & Fleet, M. E. "Sulfur species at chalcopyrite (CuFeS2) fracture surfaces," Amer. Mineral., 2004, Vol. 89, pp. 1026–1032.

[22] Harmer, S. L., Pratt, A. R., Nesbitt, H. W. & Fleet, M. E. "Reconstruction of fracture surfaces on bornite," *The Canadian Mineralogist*, Can. Mineral, 2005, 43, 1619-1630.

[23] Harmer, S. L.; Skinner, W. M.; Buckley, A. N. & Fan, L.-J. Species formed at Cuprite fracture surfaces; observations of O 1s core level shift. Surf. Sci., 2009, 603, 537–545.

[24] Wang, N.; Liu G.; Dai, H.; Ma, H.; Lin, M. Spectroscopic evidence for electrochemical effect of mercury ions on gold nanoparticles, Elsevier, Analytica Chimica Acta, 2009, 1062, 140-146.

[25] Brion, D. Etude par spectroscopy de photoelectrons de la degradation superficielle de FeS$_2$, CuFeS$_2$, ZnS et PbS a l'air et dans l'eau, Appl. Surf. Sci., 1980, 5, 133–152.

[26] Buckley, A. N.; Hamilton, I. C. & Woods, R. Investigation of the surface oxidation of bornite by linear potential sweep voltammetry and X-ray photoelectron spectroscopy, J. Appl. Electrochem., 1980, 14, 63 -74.

[27] Buckley, A. N. (1994). The application of X-ray photoelectron spectroscopy to flotation research, Colloids Surf., 93, 159 -172.

[28] Buckley, A. N.; Skinner, W. M.; Harmer, S. L.; Pring, A.; Lamb, R. N.; Fan, L.-J.; Yang, Y.-W. Examination of the proposition that Cu (II) can be required for charge neutrality in a sulfide lattice-Cu in tetrahedrites and sphalerite, Can. J. Chem., 2007, 85, 767–781.

[29] Pratt, A. Photoelectron core levels for enargite, $Cu_3AsS_4$, Surf. Interface Anal., 2004, 36, 654–657.

[30] Smart, R. S. C.; Amarantidis, J.; Skinner, W. M.; Vanier, L. L. & Grano, S. R.. Topics in Applied Physics, Vol. 85, Solid–Liquid Interfaces, edited by K. Wandelt and S. Thurgate, 2003, pp. 3–60.

[31] Smart, R. S. C., Amarantidis, J., Skinner, W., Prestidge, C. A., La Vanier, L. & Grano, S. Scanning Microsc. 1998, 12, 553–583.

[32] Buckley, A. N.; Hamilton, I. C. & Woods, R. Proceedings of the International Symposium on Electrochemistry in Mineral and Metal Processing II, edited by P. E. Richardson and R. Woods, 1998, pp.234–246.

[33] Mikhlin, Y.; Romanchenko, A. and Asanov, I. Oxidation of arsenopyrite and deposition of gold on the oxidized surfaces: a scanning probe microscopy, tunneling spectroscopy, and XPS study. Geochim. Cosmochim. Acta, 2006a, 70, 4874–4888.

[34] Murphy, R. and Strongin, D.R. Surface reactivity of pyrite and related sulfides, Elsevier, Surface Science Reports, 2009, 64 1- 45.

[35] Sanchez-Arenilla, M. and Mateo-Marti, E. Pyrite surface environment drives molecular adsorption: cystine on pyrite (100) investigated by X-ray photoemission spectroscopy and low energy electron diffraction, Phys. Chem. Chem. Phys., 2016, 18, 27219 – 27225.

[36] Okazawa, T.; Fujiwara, M.; Nishimura, T.; Akita, T.; Kohyama, M.; Kido, Y. Growth mode and electronic structure of Au nano-clusters on NiO (001) and $TiO_2$(110), Surf. Sci., 2006, 600, 1331.

[37] Chen, M.; Cai, Y.; Yan, Z.; Goodman, D.W. On the origin of the unique properties of supported Au nanoparticles, J. Am. Chem. Soc., 2006, 128, 6341.

[38] Okazawa, T.; Kohyama, M.; Kido, Y. Electronic properties of Au nano-particles supported on stoichiometric and reduced $TiO_2$ (110) substrates, Surf. Sci., 2006, 600, 4430.

[39] Acres, R.G.; Harmer, S.L; Beattie, D.A. Synchrotron XPS, NEXAFS, and ToF-SIMS studies of solution exposed chalcopyrite and heterogeneous chalcopyrite with pyrite, Elsevier Minerals Engineering, 2010, 23  928–936.

[40] Luo, M. F.; Wang, C.C.; Hu, G.-R.; Lin, W. R.; Ho, C.Y.; Lin, Y. C.; Hsu, Y. J. Active alloying of Au with Pt in nanoclusters supported on a thin film of $Al_2O_3$/NiAl (100), J. Phys. Chem. C, 2009, 113, 21054.

[41] Boyen, H.G.; Herzog, T.; Kastle, G.; Weigl, F.; Ziemann, P.; Spatz, J. P.; Moller, M.; Wahrenberg, R.; Garnier, M.G.; Oelhafen, P. X-ray photoelectron spectroscopy study on gold nanoparticles supported on diamond, Phys. Rev. B, 2002, 65.

[42] Boyen, H.G.; Kastle, G.; Weigl, F.; Ziemann, P.; Schmid, G.; Garnier, M.G.; Oelhafen, P. Chemically induced metal-to-insulator transition in Au55 clusters: effect of stabilizing ligands on the electronic properties of nanoparticles, Phys. Rev. Lett., 2001, 87, 276401.

[43] Buttner, M. and Oelhafen, P. XPS study on the evaporation of gold sub-monolayers on carbon surfaces, Surf. Sci., 2006 600, 1170.

[44] Wertheim, G.K.; Dicenzo, S.B.; Youngquist, S.E. Unit charge on supported gold clusters in photoemission final-state, Phys. Rev. Lett., 1983, 51, 2310.

[45] Hofstra, A. H.; Cline J. S. Characteristics and Models for Carlin-Type Gold Deposit, SEG Reviews, 2000, 13, 163-220.

[46] Bayari, E.E.; Foli G. and Gawu, S.K.Y. Geochemical and pathfinder elements assessment in some mineralized regolith profiles in Bole-Nangodi gold belt in north-eastern *Ghana,* Environmental Earth Sciences, 2019, 78:268.

[47] Nude P. M.; Asigri J. M; Yidana S. M; Arhin E.; Foli G.; and Kutu J. M. Identifying Pathfinder Elements for Gold in Multi-Element Soil Geochemical Data from the Wa -Lawra Belt, Northwest Ghana: A Multivariate Statistical Approach. International Journal of Geosciences, 2012, 3, 62-70.

[48] Marfo, E.; Darko, E.; Faanu, A.; Mayin, S. Study of the radiological parameters associated with small-scale mining activities at Dunkwa-on-Offin in the central region of Ghana, Radiation Protection and Environment, 2016, Vol. 39, 2.

[49] Wright, J.B.; Hastings, D.A.; Jones, W.B.; Williams, H.R. and Wright, J.B. (ed.). Geology and Mineral Resources of West Africa. London: George Allen & UNWIN. 1985, pp. 45–47.

[50] Kesse, G.O.; Foster, R.P.; (ed.): - The occurrence of gold in Ghana, in Gold '82: The Geology, Geochemistry and Genesis of Gold Deposits. Rotterdam: Geological Society of Zimbabwe, A.A. Balkema. 1984, pp. 648–650.

[51] Kim, B.J.; Cho, K. H.; Lee, S.G.; Park, C-Y.; Choi, N.C. and Lee, S. Effective Gold Recovery from Near-Surface Oxide Zone Using Reductive Microwave Roasting and Magnetic Separation, Metals, 2018, 8, 957; doi:10.3390/met8110957.

[52] Craw, D.; and Lilly, K. Gold nugget morphology and geochemical environments of nugget formation, southern New Zealand. Ore Geol. Rev., 2016, 79, 301–315.

[53] Craw, D.; MacKenzie, D.J.; Grieve, P. Supergene gold mobility in orogenic gold deposits, Otago Schist, New Zealand. NZ J. Geol. Geophys., 2015, 58, 123–136.

[54] Webster, J.G. The solubility of Au and Ag in the system Au–Ag–S–$O_2$–$H_2O$ at 25 C and 1 atm. Geochim. Cosmochim. Acta, 1986, 50, 245–255.

[55] Kratos Analytical Ltd.: library filename: "casaXPS_KratosAxis-F1s.lib".

[56] Nzulu, G.; Eklund, P. and Magnusson, M. Characterization and Identification of Au Pathfinder Minerals from an Artisanal Mine Site using X-Ray Diffraction, J. Mater. Sci. 2021, 56:7659-7669.

[57] Fuggle, J. C. and Mårtensson, N. "Core-Level Binding Energies in Metals," J. Electron Spectrosc. Relat. Phenom., 1980, 21, 275.

[58] Kishi K, Nishioka J. Interaction of Fe/Cu (100), Fe-Ni/Cu (100) and Ni/Fe/Cu (100) surfaces with $O_2$ studied by XPS. Surf Sci, 1990, 227(1-2):97–106.

[59] Lozzi L.; Passacantando M.; Picozzi P.; Santucci S.; Den Haas H. Oxidation of the Fe/Cu (100) interface. Surf Sci, 1995, 331:703–709.

[60] Mirsaleh-Kohan, N.; Bass, A.D. and Sanche, L. X-ray Photoelectron Spectroscopy Analysis of Gold Surfaces after Removal of Thiolated DNA Oligomers by Ultraviolet/Ozone Treatment, PMC, 2001, 26, 9, 6508-6514.

[61] Dochartaigh, B. É. Ó.; Smedley, P. L.; MacDonald A. M;, Darling, W. G.; and Homoncik, S. Groundwater chemistry of the Carboniferous sedimentary aquifers of the Midland Valley, British Geological Survey, Groundwater program, 2011.

[62] Casaletto, M. P.; Longo, A.; Martorana, A.; Prestianni, A. and Venezia, A. M. XPS study of supported gold catalysts: the role of Au0 and Au+d species as active sites, Surf. Interface Anal. 2006, 38, 215–218.

[63] Kaushik V.K. XPS core level spectra and Auger parameters for some silver compounds, J. Electron Spectrosc. Relät. Phenom., 1991, 56, 273.

[64] Liu Y., Jordan, R.G., and Qui, S.L. (1994) Electronic structures of ordered Ag-Mg alloys, phys Rev. 49, 7, 4478

[65] XPS interpretation of silver. Webpage: https://xpssimplified.com/elements/silver.php.

[66] Gaarenstroom, W. and Winograd, N. X-ray photoemission studies of atom implanted matrices: Cu, Ag, and Au in SiO2, J. Chem. Phys., 1979, 70, 5714.

[67] Bzowski, A.; Sham, T. K.; Watson, R. E. and Weinert, M. Electronic structure of Au and Ag overlayers on Ru (001): The behavior of the noble-metal d bands, Phys. Rev. B, 1995, 51, (15) 9980-9984.

[68] Biesinger, M.C.; Payne, B.P.; Andrew, P.; Grosvenor, A.P.; Laua, L.W.M.; Gerson, A.R.; and Smart, R. St. C. Resolving surface chemical states in XPS analysis of first row transition metals, oxides and hydroxides: Cr, Mn, Fe, Co and Ni, Applied Surface Science, 2011, 257, 2717–2730.

[69] Uhlig I.; Szargan R.; Nesbitt H.W.; and Laajalehto K. Surface states and reactivity of pyrite and marcasite, Applied Surface Science, 2001, 179, 1, 222-229.

[70] Van Der Heide H.; Hemmel R.; Van Bruggen C.F.; Haas C. X-ray photoelectron spectra of 3 d transition metal pyrites J. Solid State Chem., 1980, 33, 17.

[71] Riga J.; Tenret-Noel C.; Pireaux J.J.; Caudano R.; Verbist J.J.; and Gbillon Y. Electronic structure of rutile oxides TiO2, RuO2, IrO2, studied by X-ray photoelectron spectroscopy. *Physica Scripta*, 1977, *16*, 351-354.

[72] Lebugle A.; Axelsson U.; Nyholm R.; and Mårtensson N. Experimental L and M Core Level Binding Energies for the Metals 22Ti to 30Zn, Phys. Scr., 1981, 23, 825.

[73] Castillo R.; Koch B.; Ruiz P.; and Delmon B. Influence of the Amount of Titania on the Texture and Structureof Titania Supported on Silica J. Catal., 1996, 161,524–529.

[74] Sanjines R.; Tang H.; Berger H.; Gozzo F.; Margaritondo G., and Levy F. Electronic structure of anatase $TiO_2$ oxide, J. Appl. Phys., 1994, 75, 2945.

[75] Baker, G. Detrital heavy minerals in natural accumulates: Australasian Inst. Mining and Metallurgy, Mon. Ser., 1962, 1, 146 p.

[76] Temple, A. K. Alteration of ilmenite: Econ. Ueology, 1966, v. 61, p. 695-714.

[77] Deer, W. A.; Howie, R. A.; and Zussman, Jack. Rock-forming minerals, v. 1, Ortho- and ring silicates: New York, John Wiley and Sons, 1962a, 333 p.

[78] Behner H.; Wecker J.; Matthee T.; Samwer K. XPS study of the interface reactions between buffer layers for HTSC thin films and silicon Surf. Interface Anal., 1992, 18, 685.

[79] Anpo M.; Nakaya H.; Kodama S.; Kubokawa Y.; Domen K.; and Onishi T. photocatalysis over binary metal oxides. Enhancement of photocatalytic activities of titanium dioxide in titanium-silicon oxides. J. Phys. Chem., 1986, 90, 1633.

[80] Rogers J.D.; Sundaram V.S.; Kleiman G.G.; Castro, C.G.C.; Douglas R.A.; Peterlevitz A.C. High resolution study of the $M_{45}N_{67}N_{67}$ and $M_{45}N_{45}N_{67}$ Auger transitions in the 5d series. J. Phys. F., 1982, 12, 2097.

[81] Humbert P. An XPS and UPS photoemission study of HgO Solid State Common, 1986, 60, 21.

[82] Sekiyama, A.; Yamaguchi, J.; Higashiya, A.; Obara, M.; Sugiyama, H.; Kimura, M. Y.; Suga, S.; Imada, S.; Nekrasov, I.A.; Yabashi, M.; Tamasaku, K. and Ishikawa, T. Prominent 5d-orbital contribution to the conduction electrons in gold, *New J. Phys.,* 2010, 12**,** 043045.

[83] Zhang, J-X.; Sheong F. K. and Zhenyang L. Superatomic Ligand-Field Splitting in Ligated Gold Nanoclusters, Inorg. Chem., 2020, 59, 13, 8864–8870.

[84] Wertheim, G. K.; DiCenzo, S. B.; and Youngquist, S. E. Unit charge on supported gold clusters in photoemission final state, Phys. Rev. Lett., 1983, 51, 2310.

[85] Howard, A.; Clark, D. N. S.; Mitchell, C. E. J.; Egdell, R. G.; and Dhanak, V. R. Initial and final state effects in photoemission from Au nanoclusters on $TiO_2$ (110), Surf. Sci., 2002, 518, 210.

[86] Zhang, P. and Sham, T. K. X-Ray Studies of the Structure and Electronic behaviour of Akanethiolate-Capped Gold Nanoparticles: The Interplay of size and Surface Effects, Phys. Rev. Lett., 2003, 90, 245502.

[87] Zafeiratos, S. and Kennou, S. A study of gold ultrathin film growth on yttria-stabilized $ZrO_2$(100), Surf. Sci., 1999, 443, 238.

[88] DiCenzo, S. B.; Berry, S. D. and Hartford, E. H. Photoelectron spectroscopy of single-size Au clusters collected on a substrate. J. Phys. Rev. B: Condens. Matter Mater. Phys., 1988, 38, 8465.

[89] Roulet, H.; Mariot, J.-M.; Dufour, G. and Hague, C. F. Size dependence of the valence bands in gold clusters, J. Phys. F: Met. Phys., 1980, 10, 1025.

[90] Mason, M. G. Electronic structure of supported small metal clusters, Phys. Rev. B: Condens. Matter Mater. Phys., 1983, 27, 748.

[91] Peters S.; Peredkov S.; Neeb M.; Eberhardt W. and Al-Hada M. Size-dependent XPS spectra of small supported Au-clusters, Surf. Sci., 2013, 608, 129.

[92] Takahiro K.; Oizumi S.; Morimoto K.; Kawatsura K.; Isshiki T.; Nishio K.; Nagata S.; Yamamoto S.; Narumi K.; and Naramoto H. Appl. Surf. Sci., 2009, 256, 1061

[93] Chenakinab S. P. and Kruse N. Au 4f spin–orbit coupling effects in supported gold nanoparticles, Phys.Chem.Chem.Phys., 2016, 18, 22778.

[94] Moslemia, H. and Gharabaghi, M. A review on electrochemical behavior of pyrite in the froth flotation process, Journal of Industrial and Engineering Chemistry, 2017, 47, 1–18

[95] Corkhill, C. L. and Vaughan, D.J. Arsenopyrite oxidation – A review, Applied Geochemistry, 2009, 24(12): 2342-2361